\documentclass{aa}
\usepackage{graphicx} 
\usepackage[urw-garamond]{mathdesign}
\usepackage[T1]{fontenc}
\usepackage{txfonts} 
\usepackage{natbib}
\bibpunct{(}{)}{;}{a}{}{,} % to follow the A&A style
\usepackage[english]{babel} 
\usepackage{url} 
\usepackage{color} %
\usepackage{amsmath}
\usepackage{bm}% bold math
\usepackage[normalem]{ulem}
\usepackage[toc,page]{appendix}
\usepackage[version=3]{mhchem}
\usepackage{natbib,twoopt}
\usepackage[breaklinks=true]{hyperref} %% to avoid \citeads line fills
\bibpunct{(}{)}{;}{a}{}{,}             %% natbib format for A&A and ApJ
\usepackage{wrapfig}
\usepackage{color}
\usepackage{colortbl}
\usepackage{mathtools}
\usepackage{xcolor}
\usepackage{chemfig}
\titlerunning{Role of NH$_{3}$ binding energy in the early evolution of protostellar cores} 
\authorrunning{S. Kakkenpara Suresh et al}
\graphicspath{{./}{Figures/}}
\usepackage{float} %to recognise the command 'H'
\usepackage{adjustbox} % to control the width of boxes
\usepackage{blindtext}
\usepackage{comment}
\usepackage{caption} %package to center captions in figures/tables. instead of center in the curly braces, one can substitute with left or right
\usepackage{subcaption}
\usepackage{enumitem}
\usepackage{hyperref} %to take you to the referenced/cited item when clicked on it in the text. 
\hypersetup{
    colorlinks=true,
    linkcolor=blue,
    filecolor=blue, %coloud of any files from the system referenced in the pdf document
    citecolor=blue, %colour of the citations/references
    urlcolor=blue, %colour of the url links
    pdftitle={Overleaf Example},
    pdfpagemode=FullScreen,
    } % the settings for the hyperref package - to change the url/cited item colour
\usepackage[toc]{appendix}
\usepackage[labelfont=bf]{caption} % To make terms like Fig X.x or Table X in bold
\usepackage{multirow}
\begin{document}

   \title{Role of NH$_{3}$ binding energy in the early evolution of protostellar cores}

   \subtitle{}

   \author{S. Kakkenpara Suresh
          \inst{1,2}
          \and
          O. Sipil\"a\inst{2}
          \and
          P. Caselli\inst{2}
          \and
          F. Dulieu\inst{1}
          }

   \institute{Centre for Astrochemical Studies,
              Max Planck Institute for Extraterrestrial Physics, Giessenbachstraße 1, 85748 Garching, Germany \\
              \email{shreyaks@mpe.mpg.de}
         \and
            LERMA, CY Cergy Paris Universit\'e, Observatoire de Paris, PSL University, Sorbonne Universit\'e, CNRS, LERMA, F-95000 Cergy, France \\
             % \thanks{}
             }

   \date{}

% \abstract{}{}{}{}{} 
% 5 {} token are mandatory
 
  \abstract
  % {BE important in gas-grain interactions} leave it empty if necessary  
   {NH$_{3}$(ammonia) plays a critical role in the chemistry of star and planet formation, yet uncertainties in its binding energy (BE) values complicate accurate estimates of its abundances. Recent research suggests a multi-binding energy approach, challenging the previous single-value notion.}
   {In this work, we use different values of NH$_{3}$ binding energy to examine its effects on the NH$_{3}$ abundances and, consequently, in the early evolution of protostellar cores.}
   {Using a gas-grain chemical network, we systematically vary the values of NH$_{3}$ binding energies in a model Class 0 protostellar core and study the effects of these binding energies on the NH$_{3}$ abundances.  }
   {Our simulations indicate that abundance profiles of NH$_{3}$ are highly sensitive to the binding energy used, particularly in the warmer inner regions of the core. Higher binding energies lead to lower gas-phase NH$_{3}$ abundances, while lower values of binding energy have the opposite effect. Furthermore, this BE-dependent abundance variation of NH$_{3}$ significantly affects the formation pathways and abundances of key species such as HNC, HCN, and CN. Our tests also reveal that the size variation of the emitting region due to binding energy becomes discernible only with beam sizes of 10 arcsec or less. }
   {These findings underscore the importance of considering a range of binding energies in astrochemical models and highlight the need for higher resolution observations to better understand the subtleties of molecular cloud chemistry and star formation processes. }


   \maketitle
%
%------------------------------------------------------------------
\section{Introduction}

% \item 1. Context 
% - why NH$_{3}$   ? why multi-binding energy? what is previous research?
% \item 2. Need
% \item 3. Task
% \item 4. Object/organisation (of the document)

% shedding light on the intricate interplay between molecular energetics and chemical evolution.

% Q. What is binding energy? Insert basic definition, formula; literature;\\ 
% Q. importance of NH$_{3}$  in space, for life, and as a tracer; how and why is it important in protostellar cores (how is it's abundance related to other species, discrepencies etc)\\

Since its detection \citep{PhysRevLett.21.1701}, NH$_{3}$ has been ubiquitously observed in a variety of environments like molecular clouds \citep{irvine1987interstellar}, prestellar cores \citep{crapsi2007observing}, the galactic centre \citep{winnewisser1979ammonia}, diffuse clouds \citep{liszt2006comparative}, galaxies \citep{sandqvist2017odin,gorski2018survey}, star-forming regions \citep{feher2022ammonia},  comets \citep{poch2020ammonium}, and planet-forming disks \citep{salinas2016first}. It is one of the six major molecules found in interstellar ices in the solid form \citep{Boogert2015}. It serves as an important tracer in the interiors of dense, starless cores where common tracers like CO and CS are depleted from the gas phase onto dust grains \citep{caselli1999co,tafalla2002systematic} due to temperatures as low as $\sim$ 6K \citep{crapsi2007observing, 2007A&A...467..179P} and number densities between 10$^{4}$ and 10$^{6}$ cm$^{-3}$ \citep{keto2010dynamics}. Although NH$_{3}$ is affected by freeze-out \citep{caselli2022central}, NH$_{3}$ is selectively tracing dense material, as it is not abundant in the molecular cloud surrounding dense cores (unlike CO and CS).\\

% Under such conditions, common molecular gas tracers such as CO and CS
% are depleted from the gas phase and mainly reside on the surface
% of dust grains upon freeze-out \citep{caselli1999co, tafalla2002systematic}. \\

% Ans. From Tinacci et al 2022 "one of the most observed, ubiquitous, and studied. It is found in a gaseous form toward the Galactic Center warm molecular clouds and cores,38,39 diffuse clouds,40 massive hot cores,41 molecular outflows,42 solar-type protostars,43 cold molecular clouds,44 prestellar cores,45 and protoplanetary disks.46 NH$_{3}$ is also observed to be quite abundant in the icy mantles that envelope the interstellar dust grains in cold regions.47 ". Also take from introduction of first paper why NH$_{3}$  is important and the references therein. NH$_{3}$  traces dense cold cores where common tracers like CO freeze out (insert reference for NH$_{3}$  detected in cold cores). 

% q. Why IRAS 16293? 

% Simulations/models cannot explain observations of NH3 (is there evidence for this?, if yes, insert refernces). \textit{Or} Models using only gas-phase formation routes cannot explain the observed abundances (of NH3?). Hence, it is important to explore grain-surface formation routes to account for this. Grain surfaces act as a meeting point for species to adsorb onto, meet and react.

An important parameter that influences the grain surface abundance of NH$_{3}$ (and other species) is the binding energy. It serves as a measure of the strength of the interaction between the species and an adsorbate or grain surface. Additionally, once the species is adsorbed onto dust grains, the binding energy plays a crucial role in governing both its diffusion across and residence time on the grain surface. These factors, in turn, significantly influence the physical and chemical conditions governing the evolution of star-forming regions. For example, in the vicinity of a low-mass protostar, the binding energy determines the distance from the central protostar at which sublimation fronts emerge due to the desorption of volatile species from ice grain surfaces. Furthermore, the location of these sublimation fronts, also known as "snowlines" in protoplanetary disks, directly shapes the composition of the planets forming within protoplanetary disks. Hence, precise knowledge of the binding energies is critical for an accurate interpretation of observational data and reliable prediction of abundances through simulations. \\

% The BE is dependent on the species, the substrate, and the coverage of the species on the substrate. The binding energy plays a key role in the physical and chemical conditions that govern stellar evolution. The observed abundances of some species (insert example) cannot be explained via gas-phase formation routes alone. Hence, the exploration of gas-grain formation mechanisms is important. Grain surfaces play a crucial role in the formation of several species. They act as a meeting point for two species to be adsorbed and react (the Langmuir-Hinshelwood Mechanism). An important parameter determining the formation of a species is its binding energy of the grain (or ice) surfaces. 

Earlier studies, such as \cite{hama2013surface},  \cite{penteado2017sensitivity} and \cite{wakelam2017binding}, provided a singular binding energy value for NH$_{3}$ on different types of water ice. However, recent research has challenged this notion, suggesting that NH$_{3}$ exhibits a distribution of binding energies. Experiments conducted by \cite{he2016binding} revealed a surface coverage-dependent distribution of binding energies for NH$_{3}$ at sub-monolayer coverages. Additionally, \cite{suresh2024experimental} observed a distribution of binding energies on crystalline ice (3780K -- 4080K) and compact amorphous solid water ice (3630K -- 5280K) at monolayer coverages. Theoretical calculations by \cite{ferrero2020binding}, \cite{tinacci2022theoretical}, and \cite{Germain2022} have also supported this conclusion. The notion of a binding energy distribution is logical given that the structure of ice on dust grains is amorphous, resulting in the formation of diverse and unique adsorption sites. This idea is further substantiated by \cite{bovolenta2020high}, who obtained a Gaussian-like distribution of binding energies for hydrogen fluoride on amorphous solid water (ASW).\\

% several people have tried to determine the be of NH$_{3}$  however applications of these determined values to models of pre-stellar/proto-stellar cores are still missing. It is important to understand the real world applications of these values to have a better idea of their impact on what we observe. It will help us calculate more precisely the observed abundances or predit estimates to identify stages of evolution
% The next step is to explore the role of binding energy on the chemistry of a protostellar core. The candidate of choice is IRAS 16293-2422 - a class 0 protostar situated in the $\rho$- Ophiuchi cloud. It is located at a distance of 120 pc \citep{lombardi2008hipparcos} and has a bolometric luminosity L$_\odot$ $\sim$ 21. It exhibits "hot corino" chemistry rich in various complex organic molecules. Centimetre and mm observations have revealed the source to be a multiple-source system IRAS 16293 A and B (\cite{wootten1989duplicity}, \cite{mundy1992iras}). Source A has been further resolved into two sub-sources A1 and A2 (\cite{wootten1989duplicity}, \cite{chandler2005iras}, \cite{sadavoy2018dust}, \cite{maureira2020orbital}).  \\

In their numerical study, \cite{grassi2020novel} have illustrated that employing a binding energy distribution allows molecules to occupy higher energy binding sites, thereby increasing their residence time on the grain surface. This prolonged residence time enhances their availability to react with other molecules, even at dust temperatures that conventionally exhibit limited or no reactivity. Building upon their findings and our previous work in \cite{suresh2024experimental}, the present study evaluates the influence of incorporating a range of binding energy values for NH$_{3}$ in astrochemical models, in particular the impact of binding energy on the abundance distribution of NH$_{3}$. We also investigate the possible effect that NH$_{3}$ abundance variations may have on the distributions of other molecules chemically linked to NH$_{3}$. We do not expect the measured values of binding energy to have an impact on the chemistry in prestellar cores due to their low temperatures. Hence, in the present work, we focus our attention on protostellar cores. We focus on the very early stages of the evolution of a protostellar core, and use a model of the well-studied class 0 protostellar core, IRAS 16293-2422, as a template for our simulations. \\

% \textcolor{red}{However, for the prestellar cores, I was wondering if we consider the effect of the different binding energies on the reactive desorption efficiency of NH3 (following the new procedure described in Riedel et al. 2023). It could be interesting to see if significant difference is seen, especially in the outer envelope of the core (this was one of the points of discussion in Caselli et al. 2017). By the way, this could also be relevant for protostellar cores.}\\

This work is organised in the following way. In Section 2, we outline the physical and chemical model of the protostellar core used. Section 3 presents our results, and Section 4 delves into the implications of these findings along with potential avenues for future research. A concise summary of our work is provided in Section 5.

%--------------------------------------------------------------------
\section{Model}
\subsection{Physical Model}

\begin{figure}
    \centering
    \includegraphics[width=\hsize]{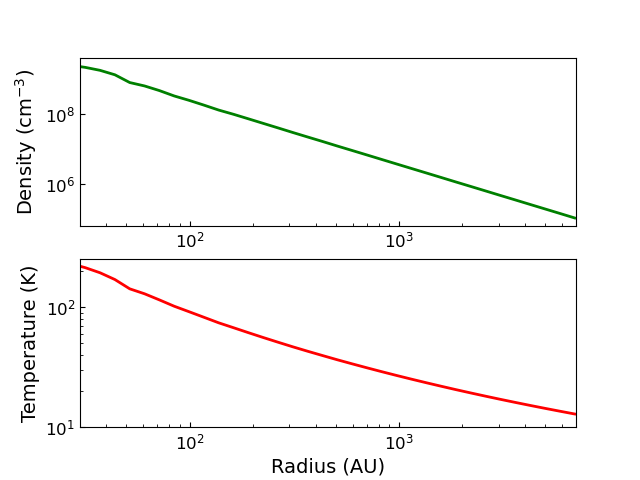}
    \caption{\small\textbf{(Top)} The H$_{2}$ number density and \textbf{(bottom)} the temperature distribution assumed to model IRAS 16293-2422.}
    \label{fig:Crimier D-T profiles}
\end{figure}

The core is characterized using the model by \cite{crimier2010solar} (Fig. \ref{fig:Crimier D-T profiles}) where the radial density profile follows as \textit{n}(H$_{2}$) $\propto$ \textit{r}$^{-1.8}$ within 6900 AU from the centre, where \textit{n}(H$_{2}$) is the H$_{2}$ number density. The source exhibits a pronounced temperature and density gradient. In this model, the temperature of the gas rises significantly towards the interior due to both gas compression during collapse and radiation emitted by the protostar. We assume that the dust temperature and the gas kinetic temperature are equivalent. The analytical approach involves dividing the core into concentric shells, and the final results are combined radially. A two-phase (gas + ice) chemical model with a dust grain radius = 0.1 $\mu$m is employed, where the entire ice layer covering the grain is available for desorption.  \\

% The model represents well single-dish observations(?): can this be used to justify the simplicity of the model? Emphasize also The extreme simplicity of the model, with there being no physical evolution between the first and second stages (no warm-up etc.). The model we use is a simple one but IRAS 16293 is a very evolved protostellar core and has been observed by MJ Maureira et al to be a multiple source system. \\

The abundance profiles are obtained in two steps. First, the initial conditions corresponding to the parent cloud are obtained by running a single-point simulation with T$_{dust}$ = T$_{gas}$ = 10K, \textit{n(H$_{2}$)} = 10$^{4}$ cm$^{-3}$, grain radius = 0.1$\mu$m, cosmic-ray ionization rate, $\zeta$ = 1.3$\times$e$^{-17}$ s$^{-1}$, and visual extinction A$_{V}$ = 10 mag.  Adsorption, desorption, and photodesorption are also considered, and the simulation is allowed to proceed until a time, t = 10$^6$ years, consistent with the previous work of \cite{brunken2014h2d+} and \cite{harju2017detection} on the source. At this stage, the abundances of all the species are extracted. These are then used as the initial abundances to trace the evolution of the protostellar core. Finally, the abundances of this core are extracted after an evolutionary time of t = 10$^4$ years. The choice of time is arbitrary; sub-structure (e.g., a protostellar disk) is expected to form within the inner regions in a timescale of 10$^{4}$ - 10$^{5}$ years which is not taken into account by the present static physical model. Hence, we chose an early time step (10$^{4}$ years) to obtain the abundances, giving an estimate of the initial chemical conditions of the forming small-scale structures.\\

%The same temperature profile is used as in \cite{2017ApJ...840...63H}. We assume that the dense core is situated/encapsulated in a low-density ambient cloud characterized by typical dark cloud conditions of n(H2) = 10^4 cm-3 and T = 10K with a thickness giving a visual extinction of 5 Av  

%What is the size of the dense cloud and that of the envelope? Do we use it?\\

%Should I insert a plot of density and temperature profile (from either Brunken or Harju)? or make my own plot reading the temperature and density from the model files in core_loop?

% \begin{table}[h]
% \centering
% \setlength{\tabcolsep}{9pt}
% \renewcommand{\arraystretch}{1.2}
% \begin{tabular}{lccc}
% \hline
% \hline
% \textbf{Model Parameters}      &&   &                                \\ \hline
% Binding Energy (K)   &&     & {3780, 4080, 5280}               \\ 
% Gas Temperature, T$_{gas}$ (K) && & \multirow{2}{*}{230 - 12}  \\ 
% Dust Temperature, T$_{gas}$ (K) &&     &                                                  \\ 
% Visual Extinction, A$_{V}$ (mag)     && & {2040 - 10}                                  \\ 
% Volume density (cm$^{-3}$)  && & $\sim$ 10$^{9}$ - 10$^{5}$          \\ \hline
% \end{tabular}
% \caption{\small Values of the physical parameters used for IRAS 16293-2422}
% \label{physical parameters}
% \end{table}

\subsection{Chemical Model}
The chemical evolution of the core is monitored by the gas-grain chemical code, \textit{pyRate}, discussed in \cite{sipila2012radial,sipila2010modelling}. The model is pseudo-time dependent, i.e., we track the chemical evolution assuming a static core. \textit{pyRate} employs the rate equation method for computing the molecular abundances. The gas-phase network is constructed based on the kida.uva.2014\footnote{https://kida.astrochem-tools.org/networks.html} network \citep{Wakelam_2015} with modifications to incorporate deuterated species and spin state chemistry \citep{sipila2015benchmarking, sipila2015spin}. Here we employ a large chemical network that contains a combined total of over 75,000 reactions in the gas phase and on grain surfaces, with which we simulate the chemistry of NH$_{3}$ and its deuterated forms. In addition, the KIDA network is also used to study the effect of proton transfer reactions on the abundances of NH$_{3}$ (Section \ref{Taquet}). \\

% The grain surface network is derived from \cite{semenov2010chemistry}.

Initially, the species are assumed to be atomic, except for H$_{2}$ and HD, and the initial ortho/para H$_{2}$ ratio is set to 10$^{-3}$ consistent with a spin temperature, $T_{spin}$ $\sim$ 20K \citep{brunken2014h2d+,crabtree2011ortho}. 
Choosing this specific value for the ratio of ortho/para H$_{2}$ corresponds to the assumption that the spin-state ratio has had the time to undergo thermalization prior to the formation of the core.  The initial abundances of the species are provided in Table \ref{Table of initial abundances}. The binding energies for NH$_{3}$ on water ice surfaces are sourced from \cite{suresh2024experimental}. Binding energies of various other species on water ice are obtained from \cite{garrod2006formation} and \cite{sipila2012radial}. The simulated abundances of NH$_{3}$   represent the combined abundances of the ortho and para forms of the molecule.\\ 

%insert formula for binding energy-desorption rate \\

\begin{table}[h]
\centering
\setlength{\tabcolsep}{9pt}
\renewcommand{\arraystretch}{1.2}
\begin{tabular}{ll}
\hline
\textbf{Species} & \textbf{Abundance} \\ \hline
H$_{2}$               & 5.00 $\times$ 10$^{-1}$            \\
He               & 9.00 $\times$ 10$^{-2}$            \\
HD               & 1.60 $\times$ 10$^{-5}$           \\
% D                & 1.00 \times 10^{-8}          \\
% H                & 1.00 \times 10^{-8}        \\
C$^{+}$               & 1.20 $\times$ 10$^{-4}$            \\
N                & 7.60 $\times$ 10$^{-5}$            \\
O                & 2.56 $\times$ 10$^{-4}$           \\
S$^{+}$               & 8.00 $\times$ 10$^{-8}$         \\
Si$^{+}$             & 8.00 $\times$ 10$^{-9}$          \\
Na$^{+}$             & 2.00 $\times$ 10$^{-9}$            \\
Mg$^{+}$              & 7.00 $\times$ 10$^{-9}$            \\
Fe$^{+}$              & 3.00 $\times$ 10$^{-9}$            \\
P$^{+}$               & 2.00 $\times$ 10$^{-10}$           \\
Cl$^{+}$             & 1.00 $\times$ 10$^{-9}$            \\ \hline
% F                & 2.00e-9            \\ \hline
\end{tabular}
\caption{\small Initial chemical abundances with respect to total H nuclei, n$_{H}$}
\label{Table of initial abundances}
\end{table}

\section{Results}
% Structuring the results section
% Begin with a brief introduction that repeats the key research question. 
% Then, the results section needs to communicate the findings of your research in a systematic manner. The section needs to be organized such that the primary research question is addressed first, then the secondary research questions. If the research addresses multiple questions, the results section must individually connect with each of the questions. This ensures clarity and minimizes confusion while reading.
% Consider representing your results visually. For example, graphs, tables, and other figures can help illustrate the findings of your paper, especially if there is a large amount of data in the results.
% Third, the results section should include a closing paragraph that clearly summarizes the
% key findings of the study. This paves the way for the discussion section of the research
% paper, wherein the results are interpreted and put in conversation with existing literature. 

% \subsection{Intro to results}
% WHAT DO I WANT TO DO? WHAT DID I DO? WHAT DID I OBSERVE?\\

% To Verify/test/analyse the impact of binding energy on the chemistry of a protostellar core, we run several models - changing only the binding energy of NH$_{3}$  between each model at each stage - and study the chemical evolution across the core. The abundances of NH$_{3}$  are extracted at a time step, t = 10e4/10e5 years. Fig. \ref{fig:abu profiles} shows the abundance profiles for NH$_{3}$  for different binding energies (insert values). \\
\subsection{Abundance Profiles}

Motivated by the experimental results of \cite{suresh2024experimental}, we investigated how variations in NH$_{3}$ binding energy impact the chemical composition of a protostellar core. For this purpose, we conducted a series of simulations where we systematically varied only the binding energy of NH$_{3}$  in each model and traced the resulting chemical evolution with time within the core. We extracted the abundances of NH$_{3}$  at t = 10$^{4}$ years representing the protostar at its early stages of formation. \\

% \subsection{Key findings:} 

\begin{figure}[h]
    \centering
    \includegraphics[width=0.45\textwidth]{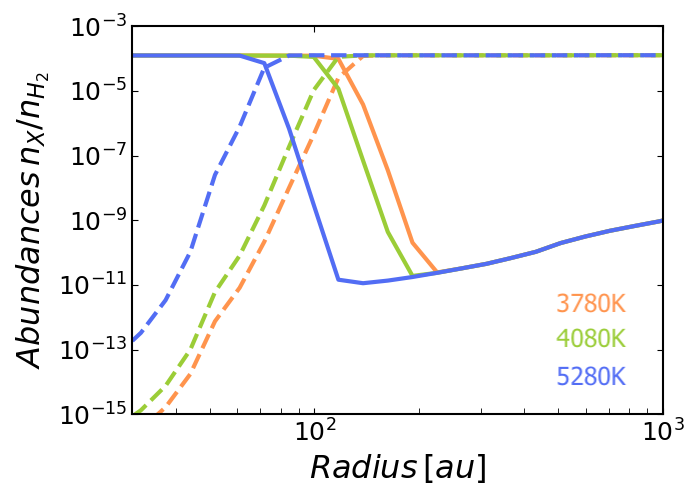}
    \caption{\small Radial abundances of NH$_{3}$  in gas phase \textit{(solid lines)} and on grain surfaces \textit{(dashed lines)} at 10$^{4}$ years. The binding energies of NH$_{3}$  used in each model (colours) are displayed in the lower right corner.}
    \label{fig:abu_profiles_1}
\end{figure}

Figure \ref{fig:abu_profiles_1} presents the abundances of NH$_{3}$ at this time in the gas-phase (solid lines) and on grain surfaces (dashed lines) with respect to the radial distance from the centre to the outer edges of the core. The gas-phase abundance of NH$_{3}$  decreases moving radially inwards toward the centre of the core before increasing rapidly by several orders of magnitude. The abundance profile shows clear differentiation based on the value of binding energy employed. The region of high NH$_{3}$ gas-phase abundance appears closer to the centre with the radius of the desorption zone varying between 150 and 300 AU with increasing binding energy value used. Investigation of the reaction rates at this time step revealed that the primary contributor influencing NH$_{3}$  gas-phase abundances is thermal desorption from dust grains.  As the binding energy increases, NH$_{3}$  remains on the grains until a higher temperature is reached at which point it acquires sufficient thermal energy for desorption. \\

\begin{figure}[h]
\centering
    \begin{subfigure}[b]{0.45\textwidth}
        \includegraphics[width = \textwidth]{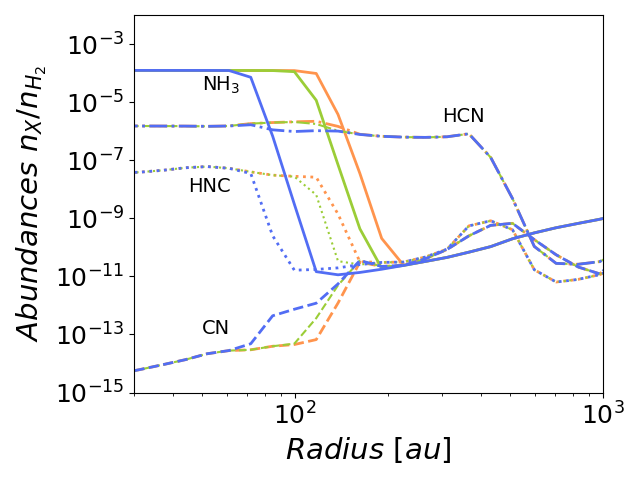}
    \caption{}
    \label{fig:gas}
    \end{subfigure}
        \begin{subfigure}[b]{0.45\textwidth}
    \includegraphics[width = \textwidth]{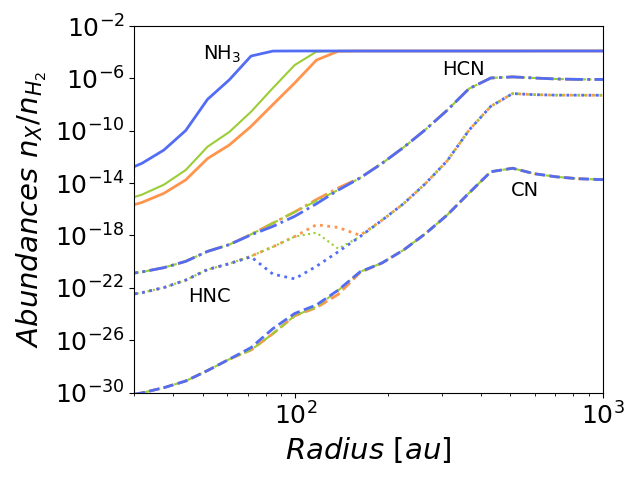}
    \caption{}
    \label{fig:grain}
    \end{subfigure}
    \caption{\small Variation in the (a) gas-phase and (b) grain abundances of NH$_{3}$, HNC, HCN and CN with binding energy. The colour scheme for the lines follows the same as in Fig. \ref{fig:abu_profiles_1}}
    \label{fig:abu_profiles_2}
\end{figure}

Notable related effects are only seen in species like HNC, CN, HCN etc (Fig. \ref{fig:abu_profiles_2}). We find that their abundance profiles vary within the same spatial zone where NH$_{3}$ abundances vary. The effects are directly tied via chemical reactions to the variations in the NH$_{3}$ abundances. This effect is further discussed in section \ref{abu profiles other species}. Furthermore, we analyzed the relative abundances of key volatile compounds found in ices, namely NH$_{3}$, CO, CO$_{2}$, CH$_{4}$, and CH$_{3}$OH with respect to H$_{2}$O. We compared our predictions with the reported values in \cite{boogert2015observations}, but direct comparisons are challenging due to differences in the sources and uncertainties related to the parameters employed for calculating ice abundances in the cited study. Further details of these comparisons are available in appendix \ref{appendix A}.\\

\subsection{Column Density maps of p-NH$_{3}$}

% % For theory, refer to subsection "Einstein coefficients" under section 3.1 in AM SHaw textbook. \\

To assess the potential observational impact of the NH$_{3}$ abundance variations, we have simulated NH$_{3}$ column density maps individually for each BE value. Here we consider p-NH$_{3}$ only so that we can compare the column density maps against the simulated emission of the (1,1) transition of this molecule (see Sect. \ref{radiative transfer}). To generate the column density maps, we employ the radial abundances of p-NH$_{3}$  obtained in the previous section as input, which are then interpolated to create a two-dimensional map of the cloud core. The abundances are convolved to a beam size of 2" for the object situated at 120 pc to simulate interferometric observations (the distance to IRAS 16293-2422, although this can be applied to other similar sources by simply changing the distance value). The radial distribution of column density, denoted as \textit{N}, is calculated using the formula: 
\begin{center}
    ${\rm N = \sum_i p \times X_i \times n(H_{2})}$ 
\end{center}

\begin{figure}
\centering
    \begin{subfigure}[b]{0.45\textwidth}
        \includegraphics[width = \textwidth]{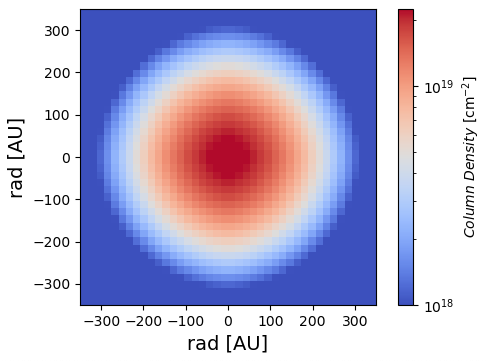}
    \caption{}
    \label{fig:colden7H3780K}
    \end{subfigure}
        \begin{subfigure}[b]{0.45\textwidth}
    \includegraphics[width = \textwidth]{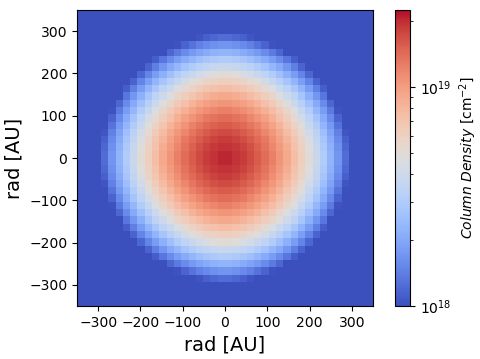}
    \caption{}
    \label{fig:colden7H4080K}
    \end{subfigure}
        \begin{subfigure}[b]{0.45\textwidth}
    \includegraphics[width = \textwidth]{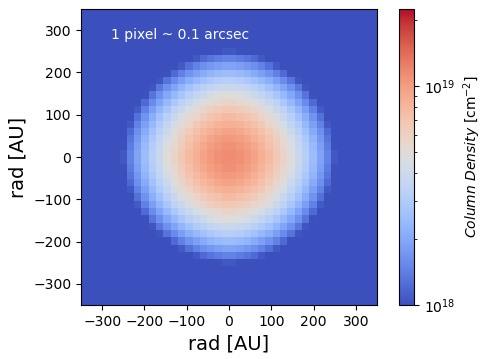}
    \caption{}
    \label{fig:colden7H5280K}
    \end{subfigure}
\caption{\small Column density map of p-NH$_{3}$  without the envelope (see text) for binding energy  (a) 3870K (b) 4080K and (c) 5280K. The angular size simulated by each pixel for each map is given in the top left corner of Fig. \ref{fig:colden7H5280K}. }
\label{fig:columndensity}
\end{figure}

\begin{figure}[h!]
\centering
    \begin{subfigure}[b]{0.5\textwidth}
        \includegraphics[width = \textwidth]{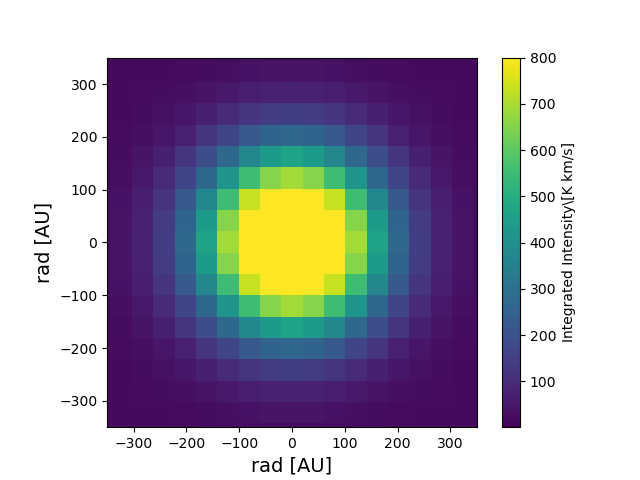}
    \caption{}
    \label{fig:intensity7H3780K}
    \end{subfigure}
        \begin{subfigure}[b]{0.5\textwidth}
    \includegraphics[width = \textwidth]{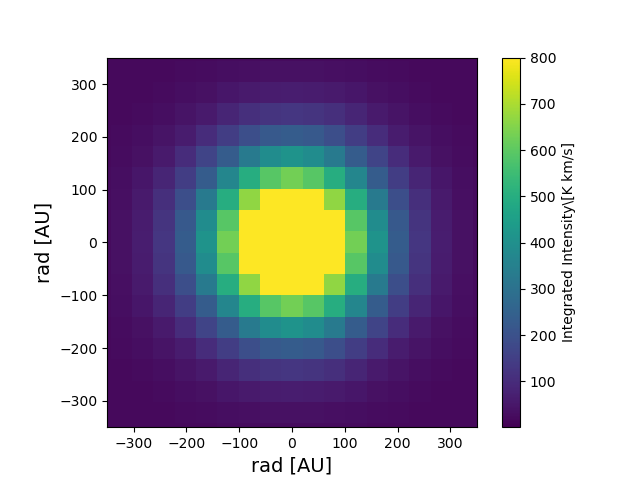}
    \caption{}
    \label{fig:intensity7H4080K}
    \end{subfigure}
        \begin{subfigure}[b]{0.5\textwidth}
    \includegraphics[width = \textwidth]{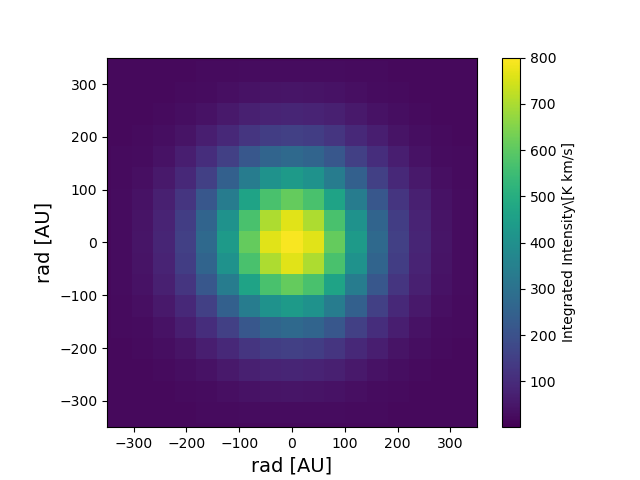}
    \caption{}
    \label{fig:intesity7H5280K}
    \end{subfigure}
\caption{\small Integrated intensity map of p-NH$_{3}$   (1,1) in the core for binding energy (a) 3870K (b) 4080K and (c) 5280K with envelope}
\label{fig:integrated intensity of pNH$_{3}$  (1,1) line with envelope}
\end{figure}

\noindent Here, \textit{$\Sigma$} represents the summation across \textit{i} elements of the core model, \textit{p} is the path length through an element \textit{i} in cm, \textit{X$_{i}$} is the abundance of the species (here, p-NH$_{3}$) in \textit{i} with respect to H$_{2}$, and \textit{n}(H$_{2}$) is the volume density of \textit{i} in mol cm$^{-3}$. Figure \ref{fig:columndensity} shows the 2D maps of column densities of p-NH$_{3}$ within a 300 AU radius where the binding energy-dependent variations are apparent. The radius of the desorption zone decreases as the binding energy increases. The column density as displayed in Fig. \ref{fig:columndensity} traces the part of the distribution that could only be observed with specific NH$_{3}$ lines, for example, the (1,1) inversion line. The column density maps also indicate the extent of desorbed NH$_{3}$ that will be available in the gas phase. \\

% \textcolor{red}{Can we also show spectra assuming a single dish observation? We could assume using the GBT telescope, so that the beam is 32 arcsec and we should not see saturated lines. If we note differences, it would be interesting to show them. }

\subsection{Radiative transfer studies of NH$_{3}$ (1,1) transition}\label{radiative transfer}

Subsequently, one-dimensional radiative transfer modelling of the 23 GHz (1,1) rotational-inversion transition of p-NH$_{3}$ is carried out to verify the observability of the binding energy-dependent variation in the radius of the desorption zone using the non-LTE radiative transfer code LOC \citep{juvela2020loc}. The first step in this analysis is to simulate the cloud that contains the protostellar core and its envelope. Given that our source is a Class 0 object, the core represents a deeply embedded source surrounded by an envelope. The methodology employed to determine NH$_{3}$ abundances within the envelope closely follows the two-step approach outlined for the core in Section 2.1. First, we obtain the abundances corresponding to the parent cloud by running a single-point simulation for t = 10$^{6}$ years under the same conditions as the parent cloud, similar to that described in Section 2.1. After extracting the NH$_{3}$ abundances at this stage, we run a second single-point simulation, this time using the specific physical conditions for the envelope. These conditions include an envelope thickness set to 0.1 pc (5 $\times$ 10$^{17}$ cm) assuming a visual extinction, A$_{V}$, = 5 mag, n(H$_{2}$) = 10$^{4}$ m$^{-3}$ and T = 10 K. The modelling is done taking into consideration the 18 hyperfine components of p-NH$_{3}$. The velocity resolution is set to $\sim$ 0.835 km/s such that it is high enough to distinguish the hyperfine components. The spectra are obtained for 1000 lines of sight across the radius of the whole object. Each spectrum is then convolved to a synthesized beam of 2". The intensity of the transition from these spectra is integrated and interpolated to create a two-dimensional intensity map (Fig. \ref{fig:integrated intensity of pNH$_{3}$  (1,1) line with envelope}) assuming spherical symmetry.\\

% The velocity resolution is set to $\sim$ 0.835 km/s such that it is high enough to distinguish the hyperfine components.

The collisional and radiative rate coefficients for the p-NH$_{3}$ line simulations were taken from the LAMDA database \citep{Schoier05}. The provided coefficients (collisional data from \cite{Danby88}) do not resolve the hyperfine structure, and hence we assumed that the coefficients for the individual hyperfine components were distributed according to LTE. To investigate the potential effect of hyperfine-resolved collisional rate coefficients on our results, we ran another series of line simulations adopting instead the set of collisional rate coefficients presented recently by \cite{Loreau23}. The hyperfine-resolved radiative transition frequencies and Einstein A coefficients were derived from data in the CDMS \citep{Endres16}. These two approaches led only to small differences in simulated lines, and in what follows we present the results of simulations carried out using the data originating in LAMDA. \\

In each model, we observe that the peak intensity is centrally concentrated and declines going outwards. The location and size of this zone are in good agreement with the simulations of the column densities (Fig. \ref{fig:columndensity}), despite optical thickness effects that were missed by calculating the column density directly from the simulated abundances. Predominantly, the intensity originates from the inner zone due to elevated NH$_{3}$  gas-phase abundance, as indicated by the abundance profiles (Fig. \ref{fig:abu_profiles_1}). Moreover, the size of the emitting region diminishes with higher binding energy, attributed to a reduction in gas-phase NH$_{3}$ concentrations associated with increasing binding energy. Similar results are obtained for radiative transfer models using the (2,2) and (3,3) lines. \\

Convolutions of the spectra for the (1,1) transitions at larger beam sizes are also conducted to evaluate the observability of this variation in the size of the emitting region. However, these effects become discernible only with beam sizes of 10 arcsec or smaller. This highlights the need for higher-resolution observations to detect such subtleties in the star-formation process. When convolved with a 6 arcsec beam, similar to in the observations by \cite{mundy1990circumstellar}, we obtain brightness temperatures lower than but within a factor of two compared to their reported value of 16.5 K for the NH$_{3}$ (1,1) emission towards the brightest regions of their source - IRAS 16293-2422. They report NH$_{3}$ emissions arising from a ring-like region of 3000-4000 AU with a resolution of 6 arcsec (960 AU). In contrast, the BE-related effects in the model used in the present work originate from a region smaller than 300 AU, indicating that higher resolution observations will be necessary to draw important conclusions regarding effects due to NH$_{3}$ binding energy towards this source. Additionally, we ran radiative transfer models assuming higher velocity resolution (0.15 km/s) for the lines based on observations by \cite{crapsi2007observing} using the Very Large Array (VLA) and found no difference in our results. Using the higher resolution requires more computational time but does not impact the results, so we decided to retain the lower velocity resolution for our models. \\

\section{Discussion}

\subsection{Influence of NH$_{3}$ binding energy on abundance profiles of chemically related species}\label{abu profiles other species}

\begin{figure}[h]
    \centering
\includegraphics[width = 0.4\textwidth]{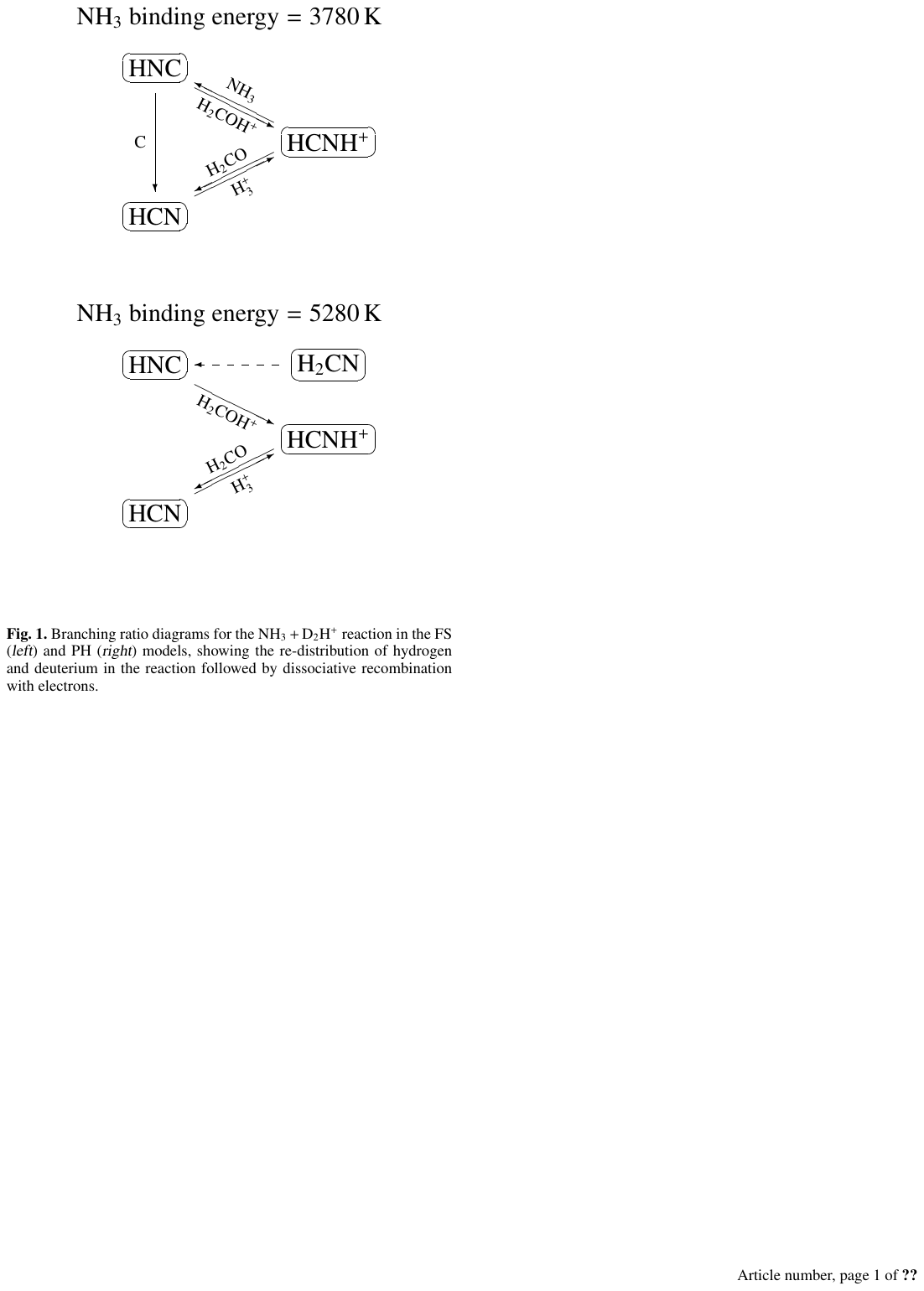}
    \caption{\small Main formation and destruction pathways of HNC as predicted by the chemical model for the two extreme binding energy (BE) values. In the lower figure (BE = 5280 K), the dashed arrow represents an alternative pathway for the formation of HNC through H-abstraction from H$_{2}$CN. This pathway becomes prominent when the gas-phase abundance of NH$_{3}$ is low, thereby inhibiting the formation via the reaction between NH$_{3}$ and HCNH$^{+}$. }
    \label{fig:reactionsHCN}
\end{figure}

Interestingly, notable effects are observed on other species and their abundance profiles within the same zone where NH$_{3}$ abundances vary. A few examples are demonstrated in Fig. \ref{fig:abu_profiles_2}. The abundance profiles of these species vary depending on the chosen binding energy of NH$_{3}$. To understand this, let us consider the example of HNC (Fig. \ref{fig:reactionsHCN}). In the inner regions of the model with binding energy = 3780K, HNC is efficiently produced through the reaction between HCNH$^{+}$ and NH$_{3}$, yielding HCN and HNC. HCNH$^{+}$ originates from HCN through the interaction with H$_{3}^{+}$ and can revert to HCN via reactions between HCNH$^{+}$ and H$_{2}$CO (and HCNH$^{+}$ + NH$_{3}$). Hence, the connection between HCN and HNC follows the sequence HCN $\rightarrow$ HCNH$^{+}$ $\rightarrow$ HNC, where the first step involves H$_{3}^{+}$, and the second step involves NH$_{3}$. 

\noindent In contrast, in the model with binding energy = 5280K, the gas phase abundance of NH$_{3}$ is low, which effectively closes off the HCNH$^{+}$ + NH$_{3}$ route. HNC is being created instead via H-abstraction from H$_{2}$CN (dashed arrow in Fig. \ref{fig:reactionsHCN}), but at approximately $\sim$10$\%$ of the rate of the NH$_{3}$ route of the previous model; HNC formation is severely inhibited by the lack of available NH$_{3}$. This can be seen in Fig. \ref{fig:abu_profiles_2} where the shape of the HNC abundance profile follows that of NH$_{3}$. Simultaneously, the rate of the reaction HCNH$^{+}$ + H$_{2}$CO $\rightarrow$ HCN + H$_{2}$COH$^{+}$ increases (as the HCNH$^{+}$ destruction channel with NH$_{3}$ is absent). These findings support the importance of further investigations into the role of NH$_{3}$ binding energy in the abundances of various species in the protostellar core. \\

% Other species HCO+, CCH, atomic C, CN, HCNH+, 

\subsection{Effect of Proton Transfer reactions on NH$_{3}$ abundances} \label{Taquet}

% A comparative analysis was performed to assess if a similar trend persists using the publicly available KIDA network \citep{Wakelam_2015}. Following the methodology outlined in section 3.1, we modeled the protostellar core using the KIDA network while varying only the NH$_{3}$ binding energy. However, the KIDA network has some limitations in that it has a smaller number of reactions ($\sim$ 7000) compared to ours and lacks a grain surface network. We address the latter by incorporating a modified version of the \cite{semenov2010chemistry} network. Moreover, KIDA does not make species separation into \textit{ortho-} and \textit{para-} forms and lacks deuteration chemistry.\\

% Despite these limitations, we find consistent trends for NH$_{3}$ where the abundance profile varies with the binding energy used, as observed in our earlier findings. Similarly, we also observe the impact of this abundance variation of NH$_{3}$ on the abundances of HCN, HNC, and CN.  

\cite{taquet2016formation} demonstrated that methanol abundances are significantly enhanced through proton transfer reactions involving NH$_{3}$, as compared to simulations lacking NH$_{3}$. Subsequently, we introduced the following reaction \\

% \begin{equation}
% \begin{align}
   \ce{ CH_{3}OH_{2}^{+} + NH_{3}   \longrightarrow CH_{3}OH_{gas}} \\
% \end{align}
% \end{equation}

% ${\rm CH_{3}OH_{2}^{+} + NH_{3}   \longrightarrow CH_{3}OH_{gas}}$

\noindent derived from the network outlined in their research, which contributes to the synthesis of dimethyl ether or methyl formate. We then ran a new chemical simulation with this reaction included, wherein only NH$_{3}$ binding energies were varied to investigate potential effects on methanol abundances. For this test, we used the KIDA network \citep{Wakelam_2015} in place of our full deuterium and spin-state containing networks, for two main reasons: 1) to check if our overall conclusions on the binding energy-dependent NH$_{3}$ desorption region remain unaffected regardless of the chemical network used (i.e., that the presence of deuterium and spin states does not affect the conclusions to a significant degree); and 2) to search for potential effects on some complex organic molecules (COMs) that are not included in our fiducial chemical network (for example, CH$_{3}$OCH$_{3}$, CH$_{3}$CH$_{2}$OH, etc). The grain-surface network used in this test is the same as our fiducial one, but with deuterium and spin states removed.\\

In the investigation of \cite{taquet2016formation}, protonation by NH$_{3}$ extended the survival time of methanol from 10$^{4}$ years to 10$^{5}$ years. In our study, methanol remained for up to 10$^{6}$ years even without their reaction integrated into the KIDA network. Upon inclusion of their reaction, our simulations validated their results, showing an increase in methanol abundances of up to two orders of magnitude for a specific binding energy value. However, this enhancement became noticeable only beyond 10$^{6}$ years. Analysis of abundances for different binding energy values at a specific time step revealed marginal variations in methanol abundances, even at 10$^{6}$ years, where the most substantial difference emerged between simulations with and without the proton-transfer reaction. The discrepancy in the evolutionary time in our model versus that of \cite{taquet2016formation} regarding when the effect on CH$_{3}$OH becomes apparent is very likely due to different simulation parameters. We have not attempted to duplicate their results. Nevertheless, the comparison indicates that proton-transfer reactions can be very important and should be explored in more detail in later simulations.\\

\section{Conclusions}

NH$_{3}$ is an important molecule widely observed in various astronomical sources and on dust grain surfaces. However, uncertainties pertaining to its binding energy values prevent an accurate determination of its abundance and role in the chemistry of star and planet formation. The binding energy is a key parameter determining the abundance and chemistry of a species in the ISM. Recent studies have challenged previous notions of a unique binding energy value for NH$_{3}$ proposing instead a multi-binding energy approach. \\

Drawing on previously established experimental work by \cite{suresh2024experimental}, we incorporated multiple  NH$_{3}$ binding energy values derived from these studies into gas-grain chemical networks to examine their impact on NH$_{3}$ abundance within a protostellar core. We conducted multiple simulations, systematically varying only the NH$_{3}$ binding energy, to discern its influence on the abundance profiles of NH$_{3}$ and other species. Our findings reveal a distinct dependence of NH$_{3}$ abundance profiles on the binding energy employed, particularly in the inner warm regions of the model. This variability extends its influence on other key species, including HCN, HNC, CN, wherein the NH$_{3}$ abundance dictates the preferred pathway for their formation. On the contrary, in proton transfer reactions involving NH$_{3}$, expected to enhance methanol formation, the abundance variation of NH$_{3}$ due to binding energy does not appear to be a significant contributing factor. \\

Simulation of column density maps for p-NH$_{3}$ reveals that the size of the desorption region diminishes as the binding energy increases. These findings align with radiative transfer studies on the (1,1) inversion line of NH$_{3}$, where we added an envelope to our physical model to examine absorption and emission effects. In these studies, we observed that the peak intensity is centrally concentrated and decreases outward. Additionally, the intensity diminishes with higher binding energy due to a reduction in gas-phase NH$_{3}$. Our results highlight the importance of considering diverse binding energies in astrochemical models, providing a refined understanding of molecular cloud chemistry and star formation processes. \\

For future work, it is crucial to refine our understanding by further exploring the impact of varied binding energies on complex organic molecules (COMs) within astrochemical models. Additionally, investigations into the spatial distribution and temporal evolution of species influenced by multi-binding energy approaches would contribute valuable insights. The present results apply at very early times in the core evolution when the substructure is still absent and, hence, probe the very early stages of forming protostellar systems. Consequently, there is a critical need to enhance the physical model, incorporating more precise representations of observed substructures within protostellar cores. This improvement is essential for providing a more accurate and detailed description of the underlying phenomena in such sources.

% Future work shall involve incorporating a physical model of the core accounting for the observed substructure. 

% How can this work be improved?
% 1. use a physical model that accounts for the observed substructure in IRAS 16293. 2. what is the cause for these observed differences in other species? 3. investigate more complex COMS

\begin{acknowledgements}
     The authors acknowledge the financial support of the Max Planck Society, CY Initiative of Excellence (grant "Investissements d'Avenir" ANR-16-IDEX-0008), the Programme National "Physique et Chimie du Milieu Interstellaire" (PCMI) of CNRS/INSU with INC/INP co-funded by CEA and CNES, a funding programme of the Region Ile de France.
\end{acknowledgements}


\newpage

\bibliographystyle{aa}
\bibliography{main}

% \printbibliography

\onecolumn
\appendix
% \begin{appendix}
\section{Relative Ice Abundances of Key Volatiles}\label{appendix A}    
    In our model, low ice abundances in the central regions of the source, attributed to the high temperatures (approximately 200K), render these ice abundance values less meaningful. As a result, in Fig. \ref{iceabuBoogert}, we present simulated abundance ratios beyond 2000 AU only, where ice abundances are higher due to temperatures dropping to around 10-20 K. Our findings for CO and CH$_{3}$OH align well with the range observed in low-mass young stellar objects (LYSOs) as described in \cite{boogert2015observations}. Although our estimates for CH$_{4}$ abundances are slightly elevated, they remain reasonably close to the values reported in the \cite{boogert2015observations}. In our model, CO$_{2}$ exhibits levels lower by two orders of magnitude, while NH$_{3}$ is overestimated by a factor of three. These discrepancies may stem from a variety of factors, such as the elemental abundances used in our model, the impact of background emissions on the observed line intensities, or missing grain-surface chemistry (in the case of CO$_{2}$). A comprehensive discussion of the cause of these variations exceeds the scope of our current study and is therefore omitted.

\begin{minipage}[t]{\textwidth}
% \begin{figure*}[b!]
\vspace{0.733cm}
    \centering
    \hspace{-1.2cm}
    \includegraphics[width=\linewidth]{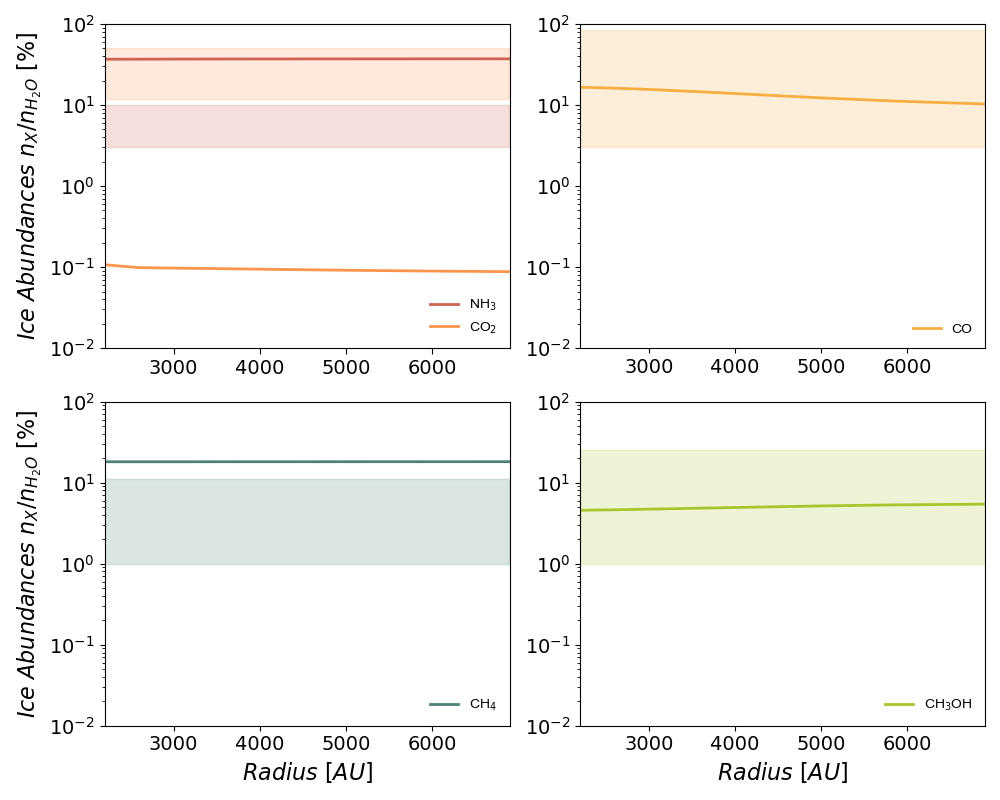}
    \captionof{figure}{\small Relative ice abundances of key volatiles with respect to water ice. The solid lines represent values obtained in this work and the shaded zones of the same colour represent the range of observed values reported in \cite{boogert2015observations} }
    \label{iceabuBoogert}
% \end{figure*}
\end{minipage}

% \end{appendix}

\end{document}